\documentclass[
reprint,
groupedaddress,
showpacs,
showkeys,
amsmath, amssymb,
aps,
pra,
]{revtex4-2}

\usepackage{xcolor}
\usepackage{graphicx}
\usepackage{dcolumn}
\usepackage{bm}
\usepackage[colorlinks=true, allcolors=blue]{hyperref}

\graphicspath{{figures/}}
\usepackage[separate-uncertainty=false]{siunitx}

\usepackage{braket}
\usepackage{xspace}
\newcommand{\na}{$^{23}$Na\xspace}

\newcommand{\down}{\ensuremath{\ket{\downarrow}}\xspace}
\newcommand{\up}{\ensuremath{\ket{\uparrow}}\xspace}

\usepackage{xstring}
\newcommand*{\aref}[1]{%
	\IfBeginWith{#1}{eq:}{Eq.~\eqref{#1}}{}
	\IfBeginWith{#1}{fig:}{Fig.~\ref{#1}}{}%
	\IfBeginWith{#1}{tab:}{Table~\ref{#1}}{}%
	\IfBeginWith{#1}{appendix:}{Appendix~\ref{#1}}{}%
	\IfBeginWith{#1}{sec:}{Section~\ref{#1}}{}%
}

\begin{document}
\title{Manipulation of an elongated internal Josephson junction of bosonic atoms}
\date{\today}

\author{A. Farolfi}
\email[]{arturo.farolfi@unitn.it}
\author{A. Zenesini}
\email[]{alessandro.zenesini@ino.cnr.it}
\author{R. Cominotti}
\author{D. Trypogeorgos}
\altaffiliation{Current address: CNR Nanotec, Institute of Nanotechnology, via Monteroni, 73100, Lecce, Italy}
\author{A. Recati}
\author{G. Lamporesi}
\author{G. Ferrari}
\affiliation{INO-CNR BEC Center, Dipartimento di Fisica, Universit\`a di Trento and TIFPA-INFN, 38123 Povo, Italy}

\begin{abstract}
We report on the experimental characterization of a spatially extended Josephson junction realized with a coherently-coupled two-spin-component Bose-Einstein condensate. The cloud is trapped in an elongated potential such that  that transverse spin excitations are frozen.
We extract the non-linear parameter with three different manipulation protocols. The outcomes are all consistent with a simple local density approximation of the spin hydrodynamics, i.e., of the so-called Bose-Josephson junction equations. 
We also identify a method to produce states with a well defined uniform magnetization.
\end{abstract}
\maketitle

\section{Introduction}

One of the macroscopic quantum effects observed in superconducting circuits and superfluid helium is the Josephson effect \cite{Josephson1962, Sato2019}, arising when two superconducting leads are coupled via tunneling effect through a thin insulating layer. 

An analogous effect has also been observed in atomic Bose-Einstein condensates (BECs). In this context, the coupling has been experimentally realized mainly in two different ways. In the first case (external coupling), two BECs are spatially separated by a thin  potential barrier that allows for tunneling \cite{Albiez2005}. In the second case (internal coupling), two internal states are Rabi-coupled by resonant radiation \cite{Zibold2010}.

As in superconducting circuits, in the most studied configurations with BECs the spatial extension does not play any relevant role and the dynamics is described by the bosonic Josephson junction (BJJ) model \cite{Smerzi1997, Steel1998} and allows to study nonlinear dynamics \cite{Albiez2005,Spagnolli2017} and non-classical states \cite{Esteve2008,Gross2010}.
 
Spatially extended systems, i.e., when the BJJ dynamics depends on the position, require a more general theoretical description, including gradients of the population imbalance and the relative phase. From the experimental point of view, double-well systems can be extended at most to 1D or 2D geometries, because the third dimension is already used to spatially separate the two quantum states. Instead, mixtures of different hyperfine states offer the possibility of studying also 3D extended systems.
So far, extended systems have not been fully investigated in experiments. Pioneering works have been reported on 1D systems studying the dynamics in the large-coupling regime \cite{Nicklas2011,Nicklas2015}, dephasing–rephasing effects \cite{Pigneur2018, Tononi2020}, local squeezing \cite{Latz2019}, and phase transition dynamics \cite{Nicklas2015b}. In this context, our group recently observed far-from-equilibrium spin dynamics dominated by the quantum-torque effect \cite{Farolfi20b}.

The control and manipulation of the full spatially extended system is  experimentally more challenging with respect to single-mode systems, because distant parts of the system can react differently depending on local properties as, for example, atomic density or external field inhomogeneities. 
Coherently-coupled systems are characterized by a flexible control thanks to the possibility of manipulating the internal state using external radio-frequency fields.
However, the techniques to prepare the system in a desired state usually act globally, hence the simultaneous control of the internal state in all spatial positions is not trivial. A strong external drive can overcome this issue, but it requires very strong uniform fields that may not be possible to implement due to technical limitations, unwanted couplings or losses to other atomic states.

In this work, we present different protocols for the manipulation of an elongated, inhomogeneous internal Josephson junction realized by coherently coupling two spin states of a Sodium BEC. The effect of the density inhomogeneity is well described within a local density approximation. 
In particular, our protocols allow us to determine the parameters of the Josephson dynamics and to prepare the whole system in an internal homogeneous state.

The paper is organised as follows:
Section \ref{Exp} introduces the experimental system and Sec.~\ref{Theo} sets the theoretical frame for an effectively 1D system. In Sec. \ref{Dim} we show that the spin dynamics in our system is one dimensional. In Sec.~\ref{Density}, we study the response of the system under a homogeneous pulse of the coupling. In Sec.~\ref{ARP} we report on an adiabatic method to produce a homogeneous state of magnetization. Finally, we present a high-accuracy characterization method for the non-linearity of the system (Sec.~\ref{plasma}).

\section{The experimental system} \label{Exp}
In our apparatus we start with a thermal cloud of $^{23}$Na atoms in a hybrid trap \cite{Colzi16, Colzi18} in the $|F,m_F\rangle=\ket{1,-1}$ state (later referred as \down), where $F$ is the total atomic angular momentum and $m_F$ its projection on the quantization axis. The atoms are then transferred into a crossed optical trap where a uniform magnetic field is applied along the $z$-axis with a Larmor frequency of \SI{913.9(1)}{\kHz}. The shot-to-shot stability \cite{Farolfi19} of the magnetic field is at the level of a few \SI{}{\micro G} over tens of minutes of continuous experimental cycling.

\begin{figure}[t]
  \centering
  \includegraphics[width=\columnwidth]{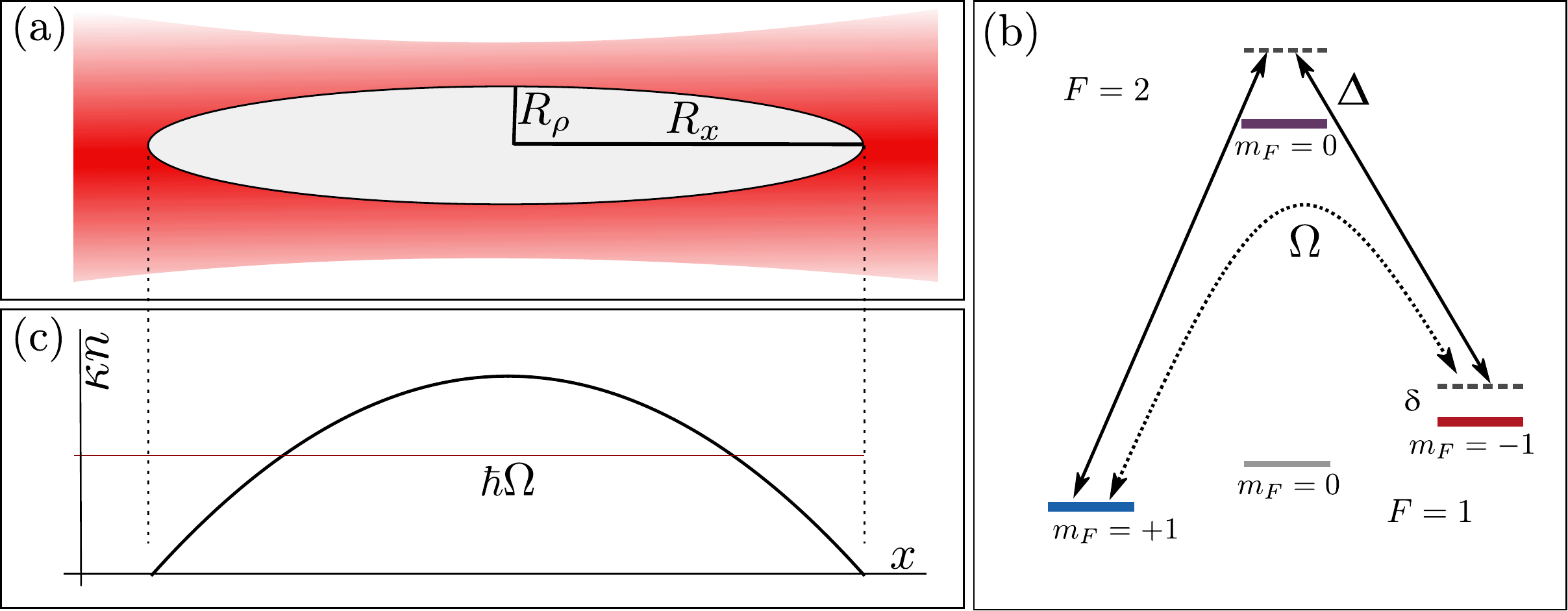}
  \caption{(a) The trapped cloud presents an elongated and cylindrically symmetric shape, with radial and axial sizes $R_\rho$ and $R_x$.
  (b) Level scheme and microwave radiations used to couple the two states $\ket{1,\pm1}$. 
  $\delta$ represents the detuning between the two-photon coupling and the $\ket{1,\pm1}$ energy difference.
  $\Delta$ is the detuning from the virtual state $\ket{2,0}$.
  (c) The non-linear strength $\kappa n$ of the cloud in the $x$ direction follows the Thomas-Fermi inverted parabola.}
  \label{fig:fig0}
\end{figure}

Evaporative cooling by lowering the depth of the optical trap leads to a BEC with up to N=$3 \times 10^6$ atoms with negligible thermal component. The atom number is known with an uncertainty of about 20\%, inferred from the calibration of the imaging system. The final trap frequencies are tuned to different values increasing the depth of the optical trap adiabatically. The final trap geometry is elongated, with axial and radial trap frequencies of $\omega_x /2 \pi \approx \SI{10}{Hz}$ and $\omega_\rho / 2 \pi$ between 500(10) and 1000(10)Hz.
The density of the sample follows the Thomas-Fermi distribution $n_{3D} = n_{3D,0}(1 - \frac{\rho^2}{R_\rho^2} -\frac{x^2}{R_x^2})$ and the radial and axial sizes are given by the Thomas-Fermi radii $R_{\rho}$ and $R_{x}$ [\aref{fig:fig0}(a)].

A two-photon Raman microwave transition to the $\ket{1,+1}$ (later referred as \up) is suddenly introduced [see Fig.\ref{fig:fig0}(b)].
The two microwave frequencies are detuned by $\Delta$ from the state $\ket{2,0}$.  The effective Rabi coupling $\Omega$ between $\down$ and $\up$ is inversely proportional to $\Delta$ and we use the latter to tune $\Omega$ while keeping the single-photon Rabi frequencies fixed to \SI{5.0(1)}{\kHz}.
The two-photon coupling can be detuned from the $\ket{1,\pm 1}$ transition by $\delta$, that we tune by varying the magnetic field.
An additional microwave radiation (\SI{20}{\kHz} blue-detuned from the $\ket{1,0} \rightarrow \ket{2,0}$ transition and with Rabi frequency of \SI{7.9(1)}{\kHz}) introduces a quadratic Zeeman shift on the $\ket{1,0}$ to suppress spin-changing collisions  \cite{Fava18}. 
The two-photon coupling and the dressing are generated by two out-of-vacuum half-dipole antennas fed by 100W amplifiers.
We routinely calibrate the magnetic field and the Rabi coupling by driving Rabi dynamics in a very dilute thermal cloud.
By driving Rabi oscillations on such a thermal cloud, we observe coherence times of \SI{370}{\ms}, presumably limited by residual collisional effects and technical noise, therefore we consider fully coherent dynamics since all the measurements are performed with less than \SI{100}{ms} of evolution time.

After applying the coherent coupling for a given time $t$, the atoms are released from the optical trap. After a short time of flight,  the states $\ket{1,\pm1}$ are separately transferred  by microwave pulses to  the stretched states $\ket{2,\pm2}$ and independently imaged by absorption imaging.

\section{Theoretical model} \label{Theo}
As mentioned in the Introduction we are interested in describing our system in terms of an inhomogeneous, elongated  BJJ. In a standard two-level approximation, it is common to use the relative population of the two states and their relative phase as the degrees of freedom of a BJJ (see, e.g., \cite{Raghavan1999}). 
However, in order to reduce the full description of our system to the one of a (local) BJJ, it is convenient  to  describe  the BEC in terms of its (position-dependent) total density $n_{3D}$ and its spin-density $\mathbf{s} = (\sqrt{n^2_{3D} - s_z^2}\cos{\phi}, \sqrt{n^2_{3D} - s_z^2}\sin{\phi}, s_z)$ on the Bloch sphere, where $s_z$ is the population difference and $\phi$ is the relative phase of the \up and \down states. The spin density has the property that $|\mathbf{s}|=n_{3D}$.
For sodium atoms, states \up and \down have equal intrastate coupling constants $g_{-1} = g_{+1} = g$ and a smaller interstate coupling constant $g_{-1, +1}$, with a positive difference $\delta g = g - g_{-1, +1}$. This leads to a full miscibility of the spin mixture \cite{Knoop2011,Bienaime2016, Fava18} and a separation of timescales between the density and spin dynamics.
Neglecting both density and spin currents, the total density is constant and the spin dynamics is described by the nonlinear precession equation \cite{Nikuni2003,Farolfi20b} 

\begin{equation}
    \dot{\mathbf{s}}({\mathbf{r}}) = \mathbf{H}(\mathbf{s}) \times \mathbf{s}({\mathbf{r}}),
    \label{eq:H_Josep}
\end{equation}
where  $\mathbf{H}(\mathbf{s}) = \left(\Omega, 0, \delta+\frac{\delta g}{\hbar}s_z\right)^T$ is the effective magnetic field.
The effective magnetic field is due to the presence of $SU(2)$ symmetry breaking terms: the homogeneous transverse microwave Rabi coupling $\Omega$, the linear detuning $\delta$ and the nonlinear detuning $\frac{\delta g}{\hbar} s_z$. The latter term arises from the difference between the intra- and interspecies interaction constants $\delta g$.

In the case of strongly-elongated cylindrically-symmetric Thomas-Fermi profile (also referred later as 1D regime, see Sec.\,\ref{Dim}), spin dynamics occurs only in the axial direction. By integrating in the radial plane, we can describe the dynamics of the spin along the axial direction $x$ introducing the 1D spin-density $\mathbf{s}(x)$, such that
\begin{equation}
   |\mathbf{s}(x)|=n(x)=n_0(1-x^2/R_x^2). 
   \label{eq:TF}
\end{equation}
The spin-density obeys the following 1D version of Eq.(\ref{eq:H_Josep}):
\begin{equation}
\dot{\mathbf{s}}(x) =\begin{pmatrix}
           \Omega \\
           0 \\
           \delta+\kappa s_z(x)
         \end{pmatrix} \times \mathbf{s}(x),
         \label{eq:H_Josep1}
\end{equation}
where the nonlinear coupling strength is 
\begin{equation}
    \kappa=\frac{5}{6\hbar}\frac{\delta g}{\pi R_\rho^2},
\end{equation}
and is related to the 3D density [see \aref{fig:fig0}(c)] through
\begin{equation}
    \kappa n_0=\frac{2}{3\hbar}  n_{3D,0} \delta g.
    \label{eq:kappa}
\end{equation}

The equations for the spin of the system,  \aref{eq:H_Josep} and \aref{eq:H_Josep1} are equivalent to a \textsl{local} version of the BJJ equations \cite{Raghavan1999}, which are written in terms of the normalized magnetization $Z(x) = s_z(x) / n$ and of the relative phase $\phi(x)$. 
In such a context, it has been realized that the BJJ equations have different dynamical regimes.
In the particular case of $\delta=0$, for $\Omega > |\kappa n|$ the dynamics resembles Rabi oscillations for any initial state. For $\Omega < |\kappa n|$, instead, a self-trapped regime characterized by a fixed-sign magnetization appears for initial states such that $\frac{Z^2}{\sqrt{1-Z^2}}>\frac{2\Omega}{kn}\cos\phi$. For $\Omega < |\kappa n|/2$, the initial states $Z = \pm 1$ are also self-trapped.
The nonlinear term in $\mathbf{H}$ is referred to as magnetic anisotropy in the context of ferromagnetism and as a capacitive term in the context of Bose-Josephson dynamics (see also below).

Equation \ref{eq:H_Josep} does not take into account density nor spin currents. The effects of these currents can be implemented by means of a full hydrodynamic description of the system \cite[Chap. 21]{Pitaevskii16} and the evolution equation becomes equivalent to the Landau-Lifshitz equation \cite{Landau1935}. For the measurements  presented here, the contribution is negligible as the applied protocols do not excite strong spin gradients.
The nonlinear term $\kappa n_0$ can be calculated from the experimental parameters (atom number and trap frequencies), but the accuracy remains poor. In Sec.~\ref{Density}, Sec.~\ref{ARP} and Sec.~\ref{plasma},  we show how we extract it from the spin dynamics in different ways.

\section{Dimensionality reduction} \label{Dim}

The dynamics of the density and of the pseudo-spin in an elongated two-component Bose-Einstein condensate can be either effectively one- or three-dimensional depending on the characteristic lengths of spin and density excitations in comparison to the radial size of the condensate. 
In an equally populated uniform sample with total density $n_{3D,0}$, the density and spin excitations are characterized by the healing length $\xi = \hbar / \sqrt{2mn_{3D,0}g}$ and by the spin healing length $\xi_s = \hbar / \sqrt{2mn_{3D,0}\delta g}$, respectively.
The ratios $R_\rho/\xi$ and $R_\rho/\xi_s$, evaluated in the center of the sample, depend on the choice of the trap parameters and the peak density $n_{3D,0}$ as follows
\begin{equation}
  \frac{R_\rho}{\xi}= \frac{2 n_{3D,0} g}{\hbar\omega_\rho }
\end{equation}
\begin{equation}
  \frac{R_\rho}{\xi_s}= \frac{2 n_{3D,0} g}{\hbar\omega_\rho }\sqrt{\frac{\delta g}{g}}.
\end{equation}
In our case, $R_{x}$ is always much larger than $\xi$ and $\xi_s$.

\begin{figure}
  \centering
  \includegraphics[width=\columnwidth]{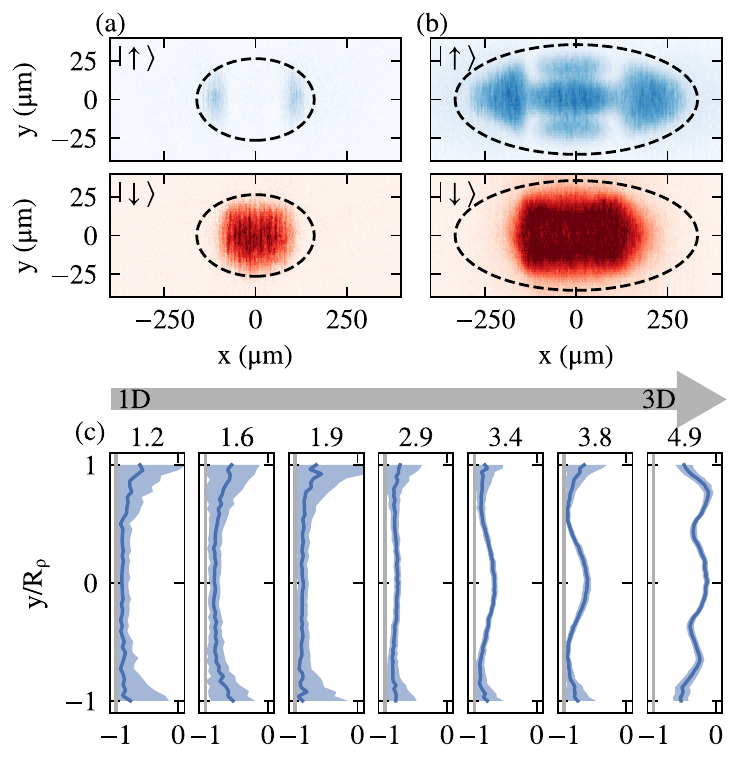}
  \caption{ Spatial distribution of the two components (\up in blue and \down in red) for an effectively 1D (a) and 3D (b) sample. $R_\rho/\xi_s$ are 1.6 and 4.9, respectively.
  (c) Magnetization along $y$ at $x=0$, integrated along $z$, after a $\pi$-pulse for different values of $R_\rho / \xi_s$ (values are reported above the plots). Confidence interval of one standard deviation is indicated as shaded region.}
  \label{fig:fig1}
\end{figure}

Since $\sqrt{\delta g/g} = 0.26$ for our sodium mixture, we can tune the experimental conditions to  effectively realize a 1D system for spin dynamics ($\xi_s \sim R_\rho$), while the total density of the sample is still well described by the Thomas-Fermi approximation ($ \xi \ll R_\rho $) and the relevant quantity characterizing the radial size is simply the 3D Thomas-Fermi radius $R_\rho$. 

The following two-step protocol is used in order to discriminate 1D spin dynamics from clouds with a 3D one.
First we tune $R_\rho/\xi_s$ by changing the final trap parameters or the total atom number. Next we apply a resonant coherent coupling pulse ($\delta=0$) to the initial condensate with all atoms in \down\,for a time $t=\pi/\Omega$. 

For the low-density regions of the cloud, the applied pulse corresponds to the well known Rabi $\pi$-pulse and one expects to observe a full population transfer into the \up\,state. In the denser part, if the non-linearity term results higher than the driving frequency $\Omega$, the population remains trapped in the \down\,state, according to BJJ dynamics. For the data in \aref{fig:fig1}, we set $\hbar\Omega \approx 0.3  n_{3D,0} \delta g$.  Evidence of a 1D regime will emerge when the low density region in the radial direction follows the denser part dynamics, remaining self-trapped to \down.

Since $R_\rho$ is comparable to our imaging resolution, we let the system expand for a short time, prior to imaging, in order to magnify the radial distribution of the population. After releasing the atoms from the trap, we let them freely expand for \SI{2}{ms} (\SI{3}{ms}) before the state \down (\up) is imaged.
Due to the different expansion time, the observed clouds have different radial dimensions. We rescale the second image along $y$ by considering that the radial size expands according to $R_\rho(t) = R_\rho(0) \sqrt{1+\omega_\rho^2t^2}$ \cite{Castin1996}.
While this relation is strictly correct only for an expanding single-component condensate, we observe that this is a good approximation also for the total density of a two-component system, even in the presence of magnetic excitations. Indeed, the large energy difference between density- and spin-excitations allows the former one to dominate the expansion of the condensate. Moreover, since the expansion times are much shorter than $1/\omega_x$, the radial expansion is ensured with negligible axial motion, allowing for direct imaging of the radial distribution of population.

Figure \ref{fig:fig1}(a) and \aref{fig:fig1}(b) highlight the differences between the 1D and 3D regime. In an effective 1D system, radial features in the magnetization are absent and the population in the center of the cloud remains in \down, as shown in  \aref{fig:fig1}(a),  for which $R_\rho/\xi_s=1.6$.  When the sample is more 3D, radial excitations lead to nonuniform radial distribution, as can be seen in \aref{fig:fig1}(b), where  $R_\rho/\xi_s=4.9$. In Fig.\ref{fig:fig1}(c) we average the density along the $x$-axis for the central 100-$\mu$m region for different values of the ratio $R_\rho/\xi_s$. Note that integration along one of the radial directions happens naturally through the absorption imaging technique. We observe that the transition between radially uniform and inhomogeneuos takes place at $R_\rho/\xi_s\approx 3$. 
For comparison, single component condensates in elongated traps, admit stable topological structures in the transverse direction for  $R_\rho/\xi > 6$ \cite{Brand2002,Mateo2014}.

In the experiments reported in the next Sections we choose $R_\rho / \xi_s = 2.5$, therefore, in the following, we consider only the 1D axial dynamics. 

\begin{figure*}
  \centering
  \includegraphics[width=2\columnwidth]{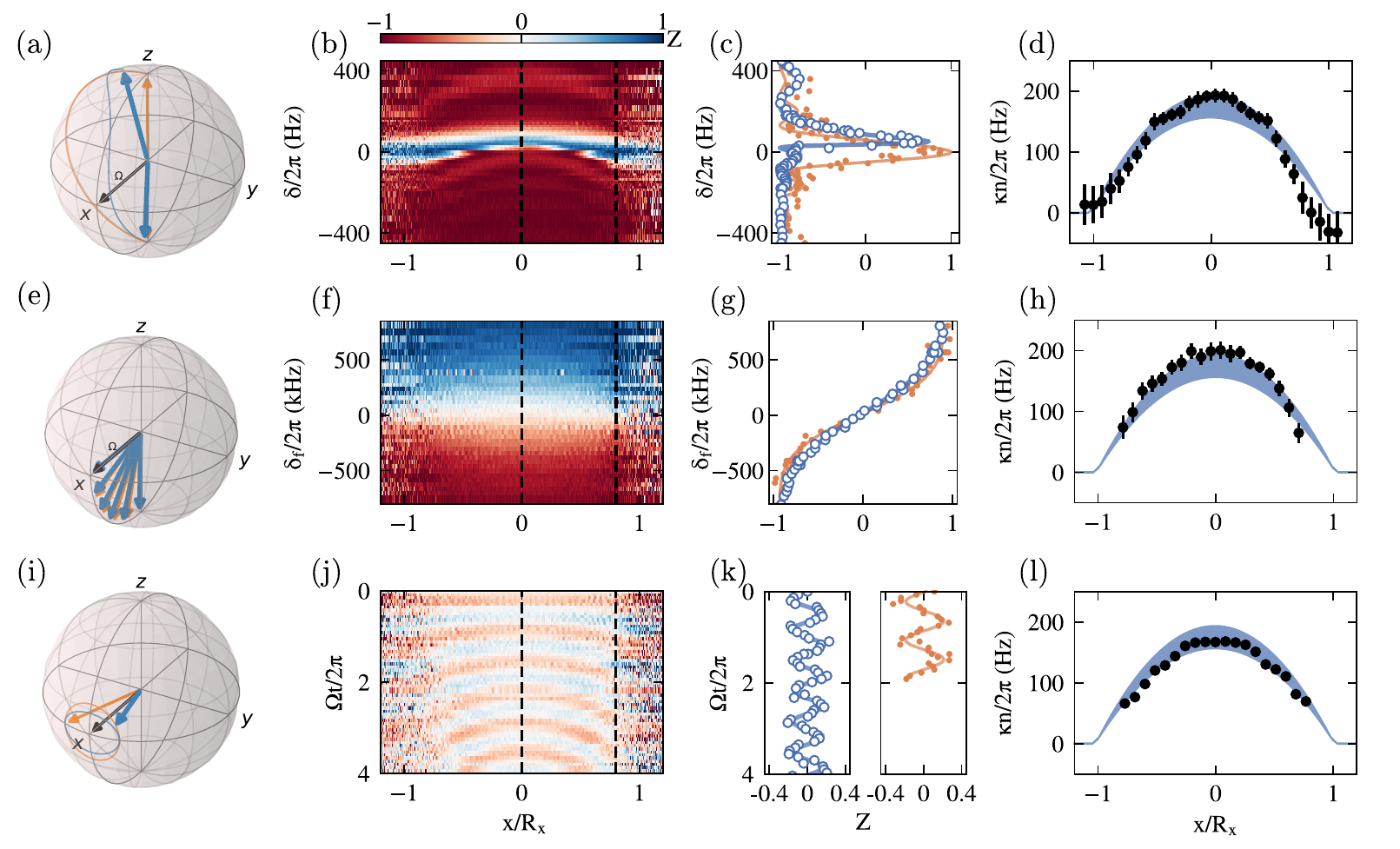}
  \caption{ 
  (a, e, i) Spin dynamics represented on the Bloch sphere in the presence (blue, thick) and in the absence (orange, thin) of the non-linear contribution for the protocols described in Sec.~\ref{Density}, Sec.~\ref{ARP} and Sec.~\ref{plasma}, respectively.
  Local magnetization for the different procedures as a function of the detuning $\delta$ (b), the final detuning $\delta_f$ (f) and time (j). The center of the cloud is at $x=0$ and extends to the Thomas-Fermi radius $R_x$. The dashed lines indicate $x = 0$ and $x = 0.8 R_x$.
  (c, g, k) Vertical cuts along the dashed lines in (b,f,j) for high- (blue circles, $x\approx 0$) and low-density (orange dots, $x= 0.8 R_x$) regions. Thick and thin lines are fits to the data for high- and low-density regions, respectively. 
  In (c), lines correspond to the solution of Eq. \ref{eq:H_Josep} with density as a fitting parameter. Significant asymmetry of the resonance peak is observed in the high-density region.
  In (g), the line is a sigmoidal function fitted to the data.
  In (k), the line is a sinusoidal function fitted to the data.
  (d, h, l) Local nonlinear parameter at different positions (points) extracted from each protocol. 
  The shaded area refers to the prediction of $\kappa n(x)$ obtained from the trap frequencies and atom number. 
  }
  \label{fig:fig2}
\end{figure*}

\section{Density-dependent shift} \label{Density}
In dense atomic clouds, transitions between energy levels are modified by the presence of interactions, whose effects can be introduced by means of  mean-field corrections. These are commonly known as collisional shifts and have great importance in metrology \cite{Harber2002}. In a Josephson system, collisional shifts are dominant when the nonlinear mean field contributions are of the same order of magnitude of (or larger than) the linear coupling strength.

Starting from a fully polarized sample in \down, a Rabi pulse with $t=\pi/\Omega$ and $\Omega = 2 \pi \times \SI{68.5(5)}{Hz}$ is applied to transfer part of the population to \up. Depending on the (global) detuning and on the (local) nonlinear contribution, the final magnetization will locally change [see \aref{fig:fig2}(a-b)]. 
The measurement is repeated for different values of the detuning $\delta$ of the coupling from the transition frequency and the final magnetization is plotted in \aref{fig:fig2}(b). 

On the thermal tails of the cloud the density is low enough that it can be considered as a pure two-level system (deep Rabi regime). In this case, the amount of transferred population depends on the detuning $\delta$ according to the commonly known sinc-like spectroscopic curve [orange data and curve in \aref{fig:fig2}(c)]. When the nonlinear term is no longer negligible compared to $\Omega$, the dynamics follows the Josephson equations [blue data and curve in \aref{fig:fig2}(c)]. The spectroscopic curve becomes asymmetric with a shifted peak. The direction and magnitude of the shift depend on the sign and magnitude of $\kappa n$, respectively.

We fit the data at different position $x$ with the numerical solution of the Josephson equation by having only $\kappa n$ as a free parameter. Figure  \ref{fig:fig2}(d) shows the local value of $\kappa n(x)$ where the error bars include statistical error on the fit of the spectroscopy data (due to shot-to-shot magnetic field and atom number fluctuation) and systematic uncertainties coming from the determination of $\delta=0$ ($\approx 10$ Hz).
With this method we obtain $\kappa n_0/2\pi=192(11)$ Hz.
The light blue area shown in \aref{fig:fig2}(d,h,l) refers to the prediction of $\kappa n(x)$ obtained from \aref{eq:kappa}, the trap frequencies and atom number being averaged over the full data acquisition. The expected value is $\kappa n_0/2\pi=173(20)$ Hz, and the main source of uncertainty is related to the determination of the atom number.

\section{Density-dependent Adiabatic Rapid Passage} \label{ARP}
Different proposals in the field of nonlinear spin-waves \cite{Qu2016, Qu2017}, quantum computation and squeezing require that the full cloud must be prepared in a single state, with a uniform magnetization $Z=0$.
For this task, the procedure presented in the previous Section can be used only if the regime $\Omega \gg \kappa n$ is experimentally reachable. In the case of $\Omega\sim \kappa n$,
the magnetization of the cloud after a pulse of duration $t=\pi/\Omega$ is not uniform as only some regions of the cloud with a certain non-linearity will be transferred for a fixed $\delta$, as \aref{fig:fig2}(b) clearly shows. A different approach is based on the Adiabatic Rapid Passage (ARP). This can be used, for instance, to generate number-squeezed states \cite{Steel1998}.

In the ARP, the coupling is applied to a polarized state with an initially large detuning, so that the system is in the state of minimum energy. The detuning is adiabatically swept to a final value $\delta_f$ close to resonance. 
During the ramp, the local magnetization  and $\delta$ are connected through the  following relation \cite{Steel1998}
\begin{equation}
\delta = \Omega \frac{Z}{\sqrt{1-Z^2}}+ \kappa n Z,
\label{eq:delta}
\end{equation}
while $\phi=0$ during the whole passage. 

Note that, far in the Rabi regime, the magnetization depends only on $\Omega/\delta$, while in the Josephson regime, an additional density-dependent term is present.
At the beginning of the ramp, all parts of the cloud are close to the south pole of the Bloch sphere. 
Due to the inhomogeneous nonlinear interaction, the magnetization has a position-dependent evolution. 
However, if $\delta$ is adiabatically reduced to zero, at the end of the ARP, the whole system will reach $Z=0$  simultaneously, independent of the value of the local nonlinear parameter, as sketched in \aref{fig:fig2}(e).

In our experiment, we start from a polarized sample in \down, turn on a coupling with $\Omega = 2 \pi \times \SI{273(1)}{Hz}$ with an initial detuning $\delta \approx 2 \pi \times \SI{3}{kHz}$.
For experimental convenience, and taking advantage from the dependence of $\delta$ on the magnetic field $B$, the sweep of the detuning is performed by keeping constant microwave frequencies and by varying the strength of the magnetic field  in \SI{50}{ms} with a nonlinear ramp. The ramp is stopped to a variable final $\delta_f$ and in \aref{fig:fig2}(f) we plot the magnetization of the sample as a function of the coordinate $x$ and $\delta_f$.

The magnetization at $\delta=0$ of the ARP procedure is less sensitive to magnetic field fluctuations, since, expanding \aref{eq:delta} near $Z=0$, one gets 
\begin{equation}
    \frac{\partial Z}{\partial \delta} = \frac{1}{\Omega + \kappa n}
    \label{eq:dZddelta}
\end{equation}
that is lowered by the nonlinear term.
Figure \ref{fig:fig2}(g) shows how the final value of the magnetization is sensitive to the final detuning, with a smaller sensitivity in the central part of the system (blue points) rather than at the edges (orange).

Remarkably this method allows for a clean preparation of the extended system in a uniform $Z=0$ state, at the expected value $\delta_f=0$, thanks to the symmetric interaction constants of \na. This result is not trivial since the magnetization varies indeed with a different velocity for each spatial coordinates. However the symmetric dynamics on the Bloch sphere leads the magnetization to reach zero at the same time for the whole cloud.
Note that the efficiency of the full rotation is increased by the nonlinear term.

By fitting the dynamics of the magnetization for each position $x$ with a sigmoidal function, we can extract the slope of the magnetization as a function of $\delta$ and hence $\kappa n$ applying \aref{eq:dZddelta} [see Fig. \ref{fig:fig2}(h)]. With such a procedure, we obtain $\kappa n_0/2\pi=200(15)$ Hz.
The error bars include statistical error on the fit and systematic uncertainties coming from the imaging procedure (uncertainty on the state population), and from a non-perfect adiabaticity of the process. Systematic contributions strongly enhance the uncertainty on the value of $\kappa n$ compared to the one obtained in Sec. \ref{Density}.

\section{Plasma oscillations} 
\label{plasma}

In the presence of coherent coupling and at $\delta = 0$, the ground-state of the system is uniformly $Z=0$, $\phi = 0$. For small deviations near the ground-state, the Josephson dynamics predicts small oscillations around $Z=0$ and $\phi=0$, which are known as plasma oscillations [see Fig.\,\ref{fig:fig2}(i)]. Their frequency follows
\begin{equation}
\omega_p = \sqrt{\Omega (\Omega + \kappa n)},
\label{eq:plasma}
\end{equation}
allowing to determine $\kappa n$ from independent measurements of $\Omega$ and $\omega_p$.

\begin{figure}[b]
  \centering
  \includegraphics[width=\columnwidth]{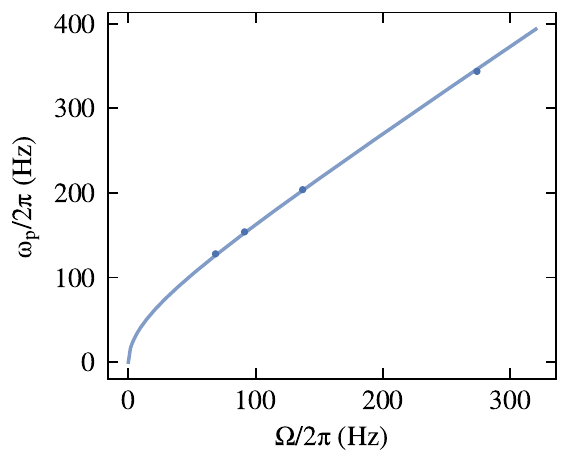}
  \caption{
  Observed oscillation frequency $\omega_p$ as function of the Rabi frequency (points). Error bars are smaller than marker size. The line is a fit of Eq.~\ref{eq:plasma} with $\kappa n$ as free parameter, yielding the value $\kappa n_0/2\pi=164(3)$.
  }
  \label{fig:fig4}
\end{figure}

The sample is prepared in $Z=0, \phi = 0$ with the previously described ARP procedure. Then, the phase of the coupling is suddenly modified from $\phi = 0$ to $\phi = 0.1\pi$, starting the oscillatory dynamics.
We extract the frequency of oscillation $\omega_p$ by fitting a sinusoid to the local magnetization. According to \aref{eq:plasma},  we determine $\kappa n$ for different $x$-position. For each fit, we determine the initial guess for the frequency by determining the peak in the Fourier-transform of the data.
In this case we obtain $\kappa n_0/2\pi=161(3)$ Hz
at the center of the cloud [\aref{fig:fig2}(k), blue points and line]. In the low-density regions of the sample the noise is larger due to the low atom number, however the observed dynamics is compatible with the independently calibrated Rabi frequency [\aref{fig:fig2}(k), orange line].
The high precision of the determination of $\kappa n(x)$ from plasma oscillations compared to previous methods is twofold. At first, fluctuations on the magnetic field, which enter as an uncertainty on $\delta$, result in a uncertainty on $\kappa n$ below 1\%. Secondly, uncertainties on the observed magnetization affect the amplitude of the oscillation, but only poorly its frequency.

We repeat the procedure for different $\Omega$. After preparation of the sample in the $Z=0, \phi = 0$ state, the detuning $\Delta$ is suddenly modified, changing the Rabi frequency. The phase is changed to $\phi = 0.1\pi$ as well.
We extract the oscillation frequency at the center of the cloud as a function of Rabi frequency \aref{fig:fig4}, and by fitting \aref{eq:plasma} on the data we determine $\kappa n$ over a range of Rabi frequencies with low statistical uncertainties.

\section{conclusions and outlook}
We have characterized the properties of an elongated Josephson junction based on two coherently coupled atomic spin states of \na. After finding the regime where the dynamics is 1D-like, we demonstrate the capability to calibrate the nonlinear term of the BJJ dynamics with different protocols. We adiabatically manipulate the internal state on the Bloch sphere to produce a uniform magnetization sample.
Additionally to the presented ARP procedure, future investigation can be focused on the search for shortcuts to adiabaticity, based on different ramps of the driving detuning and amplitude, in order to decrease losses and decoherence of the system during the state preparation \cite{Odelin2019}. 

The full control of the quantum state of an elongated Josephson junction represents a cornerstone to future investigations in the field of nonlinear dynamics and towards new metrological tools. 
The system can be driven to points of the Bloch sphere that are far from the equilibrium position, but present locally different evolution due to the non-uniform nonlinearity, leading to localized and propagating instability \cite{Bernier2014, Farolfi20b}.
In an elongated cloud, the interplay between a spatially non-uniform squeezing and the long-range entanglement requires further theoretical and experimental investigations, with particular focus on local and global correlations \cite{Latz2019}.

While the presented results focus on the dynamics in a one-dimensional system, the possibility of tuning the effective dimensionality of the system could allow the experimental investigation of topological excitations in the transverse directions, such as domain walls and vortices \cite{Son2002, Tylutki2016, Calderaro2017}.

\section{acknowledgements}
We thank I. Carusotto and P. Hauke for fruitful discussions. We acknowledge fundings from INFN through the FISH project, from the European Union’s Horizon 2020 Programme through the NAQUAS project of QuantERA ERA-NET Cofund in Quantum Technologies (Grant Agreement No. 731473), from Italian MIUR under the PRIN2017 project CEnTraL (Protocol Number 20172H2SC4) and from Provincia Autonoma di Trento. We thank the BEC Center in Trento, the Q@TN initiative and QuTip.

\end{document}